\documentclass[a4paper]{article}

\usepackage{INTERSPEECH2021}

\title{F-T-LSTM based Complex Network for Joint Acoustic Echo Cancellation and Speech Enhancement}
\name{Shimin Zhang, Yuxiang Kong, Shubo Lv, Yanxin Hu, Lei Xie$^*$\thanks{$^*$ Lei Xie is corresponding author.
}}
\address{
	Audio, Speech and Language Processing Group (ASLP@NPU), School of Computer Science 
	Northwestern Polytechnical University, Xi'an, China}
\email{\{shmzhang, yxkong, shblv, yxhu\}@npu-aslp.org, lxie@nwpu.edu.cn}

\begin{document}

\maketitle
\begin{abstract}
  With the increasing demand for audio communication and online conference, ensuring the robustness of Acoustic Echo Cancellation (AEC) under the complicated acoustic scenario including noise, reverberation and nonlinear distortion has become a top issue. 
  Although there have been some traditional methods that consider nonlinear distortion, they are still inefficient for echo suppression and the performance will be attenuated when noise is present.
  In this paper, we present a real-time AEC approach using complex neural network to better modeling the important phase information and frequency-time-LSTMs (F-T-LSTM), which scan both frequency and time axis, for better temporal modeling.
  Moreover, we utilize modified SI-SNR as cost function to make the model to have better echo cancellation and noise suppression (NS) performance.
  With only 1.4M parameters, the proposed approach outperforms the AEC-challenge baseline by 0.27 in terms of Mean Opinion Score (MOS).
\end{abstract}
\noindent\textbf{Index Terms}: Acoustic echo cancellation, complex network, nonlinear distortion, noise suppression

\section{Introduction}

Acoustic echo is generated in a full-duplex voice communication system, where a far-end user receives a modified version of his/her own voice due to the acoustic coupling between a loudspeaker and a microphone at near-end point. Acoustic echo cancellation (AEC) aims to eliminate the echo from the microphone signal while minimizing the distortion of the near-end speaker's speech. 

Traditional digital signal processing (DSP) based AEC works by estimating the acoustic echo path with an adaptive filter~\cite{benesty2007springer, 1163949, 103078}. But in practical applications, their performance may heavily degrade due to issues such as echo path change, background noise and nonlinear distortion.

Background noise is inevitable in a real full-duplex voice communication system. However, traditional speech enhancement methods, combined with AEC~\cite{2002A}, are not robust enough to such interference especially the non-stationary noise. 
Nonlinear distortion commonly caused by low-quality speakers, overpowered amplifiers and poorly designed enclosures; even modest nonlinear distortion can degrade the performance of linear AEC models considerably~\cite{4517596}.
In general, post-filter methods~\cite{hansler2005acoustic, turbin1997comparison, boll1979suppression} are further used in traditional AEC, but these methods are still inefficient for echo suppression.

Recent advances in deep learning have shown great potential in acoustic echo cancellation due to its strong non-linear modeling ability.
There are some methods that combine traditional signal processing with neural networks to deal with the AEC task. Ma et al.~\cite{ma2020acoustic} used adaptive filter processing linear echo as well as a lightweight LSTM structure for residual non-linear echo cancellation. Fazel et al.~\cite{fazel2020cad} designed a deep contextual-attention module with frequency domain NLMS to adaptively estimate features of the near-end speech. Wang et al.~\cite{wang2021weighted} and Valin et al.~\cite{valin2021lowcomplexity} have also achieved competitive results in the recent AEC-challenge~\cite{sridhar2020icassp}.
Zhang and Wang~\cite{zhang2018deep} formulated AEC as a supervised speech separation problem, where a bidirectional long-short term memory (BLSTM) network was adopted to predict a mask for magnitude of microphone signal. After that, many AEC algorithms based on speech enhancement/separation network have been proposed. West-hausen et al.~\cite{westhausen2020acoustic} extended DTLN~\cite{westhausen2020dual} by concatinating the far-end signal as additional information. Chen et al.~\cite{chen2020nonlinear} proposed a residual echo suppression (RES) method with convolution network based on the modification of ConvTasNet~\cite{Luo_2019}, and Kim et al.~\cite{kim2020attention} proposed an auxiliary encoder and an attention network based on Wave-U-Net~\cite{stoller2018wave} to effectively suppressed the echo.

\begin{figure}
\centering
    \includegraphics[width=0.35\textwidth]{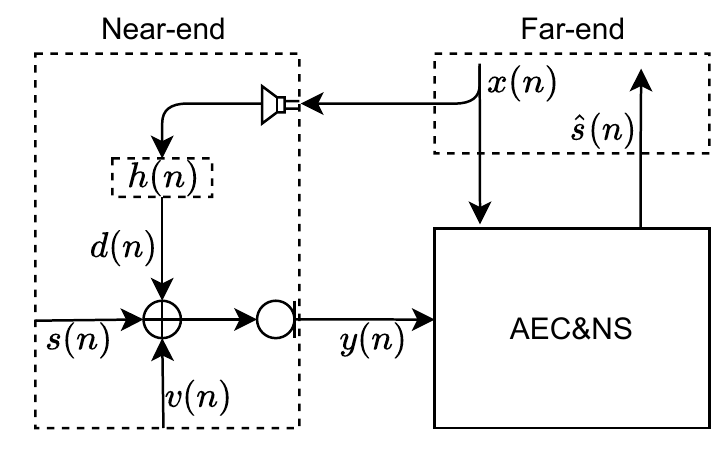}
    \vspace{-1pt}
    \caption{Diagram of an acoustic echo scenario.}
    \label{fig:aec_flow}
\vspace{-20pt}
\end{figure}

Recent studies~\cite{choi2018phase, hu2020dccrn} in speech enhancement have shown significant benefit of using a complex network, which handles magnitude and phase simultaneously, leading to superior performance in speech enhancement. Compared with a real-valued network, a complex network can even achieve better performance with much smaller size of parameters~\cite{hu2020dccrn}. The superior performance is mainly attributed to the effective use of the phase information.
Moreover, complex domain based methods have achieved overall better subjective listening performance in the Deep Noise Suppression (DNS) Challenge~\cite{sridhar2020icassp}.

In this paper, inspired by the recent advances in complex network, we address the AEC task by adopting a complex encoder-decoder structured network. To the best of our knowledge, this is the first work that adopts complex network in the AEC task. Specifically, we use complex Conv2d layers and complex Transposed-Conv2d layers as encoder and decoder respectively to model the complex spectra from both far-end and near-end signal, and complex LSTM layers as the mask estimation module.
Inspired by F-T-LSTM~\cite{li2015lstm}, we perform recurrence on frequency axis of high-dimensional features extracted by the encoder. The bi-directional F-LSTM on frequency axis allows the network to learn better the relationship between frequency bands, and the subsequent T-LSTM scans the time axis, aiming to remove the echo signal further.  We also adopt segmented Si-SNR as the cost function of our network. With only 1.4M parameters, the proposed approach outperforms the AEC-challenge baseline by 0.27 in terms of Mean Opinion Score (MOS).

\section{Proposed Method}
\subsection{Problem formulation}
\label{sec:method}
We illustrate the signal model of acoustic echo cancellation in Fig.~\ref{fig:aec_flow}. The microphone signal $y(n)$ consists of near-end speech $s(n)$, acoustic echo $d(n)$ and background noise $v(n)$:
\setlength\abovedisplayskip{2.0pt}
\setlength\belowdisplayskip{2.0pt}
\begin{equation}
    y(n) = s(n) + d(n) + v(n),
\end{equation}
where $n$ refers to time sample indexes. $d(n)$ is obtained by the far-end signal $x(n)$ as illustrated in Fig.~\ref{fig:aec_flow} and it also may have nonlinear distortion caused by speakers. $h(n)$ denotes acoustic echo path. The acoustic echo cancellation task is to separate $s(n)$ apart from $y(n)$, on the premise that $x(n)$ is known.

\begin{figure}[h]
\centering
\includegraphics[width=1.0\linewidth]{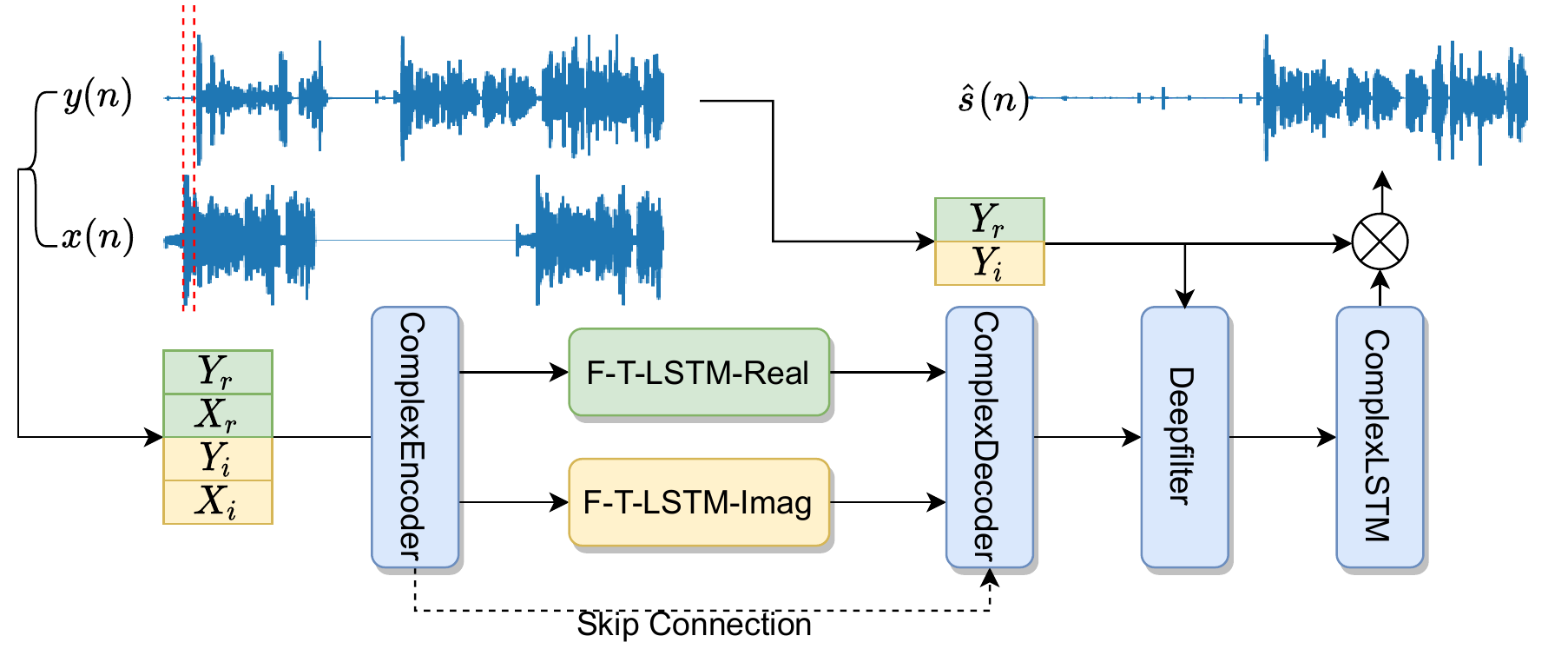}
\vspace{-15pt}
\caption{System flowchart of proposed network. (A) The red dotted area shows the time delay of between $y(n)$ and $x(n)$. (B) F-T-LSTM-real and F-T-LSTM-imag are used to model the real and imaginary parts of the high-dimensional complex feature, respectively. (C) $y(n)$ and $x(n)$ are converted to $Y$, $X$ through STFT respectively. The estimated signal $\hat s(n)$ is reconstruct through inverse-STFT.}
  \label{fig:structure}
\vspace{-20pt}
\end{figure}

\subsection{Architecture}
\label{ssec:F-T-LSTM}
As illustrated in Fig.~\ref{fig:structure}, our deep complex AEC network consists of three modules: \emph{complex encoder-decoder network}, \emph{F-T-LSTM} and \emph{complex LSTM}.

For a sequential input ${w} \in \mathbb{R}^{2 \times N}$, where $N$ is the number of audio sampling points and $2$ denotes two signals -- $y(n)$ stacks with $x(n)$. Performing STFT on the input signal ${w}$, we obtain the complex spectra ${W} = {W_r} + j{W_i}, {W} \in \mathbb{R}^{4 \times T \times F}$, where input complex matrix ${W_r}$ and ${W_i}$ represent respectively the real and imaginary part of ${W}$ with same tensor dimension $\mathbb{R}^{2 \times T \times F}$. $T$ denotes the frame number and $F$ denotes the frequency dimension after STFT. Complex convolutional/deconvolutional filter ${K}$ is defined as ${K} ={K}_r+j{K}_i$, where the real-valued matrices ${K}_r$ and ${K}_i$ represent the real and imaginary part of a complex kernel, respectively. The complex operation ${W} \circledast {K}$ defined as:
{
\setlength\abovedisplayskip{2.0pt}
\setlength\belowdisplayskip{2.0pt}
\begin{equation}
{H}=\left({K}_{r}*{W}_{r}-{K}_{i} * {W}_{i}\right)+j\left({K}_{r} * {W}_{i}+{K}_{i} * {W}_{r}\right).
\end{equation}
}
$H = H_r + jH_i, H\in \mathbb{R}^{C \times M \times T}$, $H_r$ and $H_i$ $\in \mathbb{R}^{C\times N \times T}$, $C$ denotes the output channel and $M$ denotes the frequency dimension change after convolution/deconvolution, $N = M/2$. 

The F-T-LSTM module for the real spectra can be described as follows (imaginary spectra are the same):
{
\setlength\abovedisplayskip{2.0pt}
\setlength\belowdisplayskip{2.0pt}
\begin{equation}
\begin{aligned}
&\text {F-LSTM:}\left\{\begin{array}{l}
U=\left[f\left(H_{r}^{\text {reshape }}[:, i,:]\right), i=1, \ldots, M\right] \\
V=H_{r}+U^{\text {reshape }}
\end{array},\right.\\
&\text {T-LSTM:}\left\{\begin{array}{l}
Z=\left[h\left(V^{\text {reshape }}[:, i,:]\right), i=1, \ldots, T\right] \\
Z_{\text {out }}=V+Z^{\text {reshape }}
\end{array},\right.
\end{aligned}
\end{equation}
}
where $H_{r}^{\text{reshape}}$ and $U$ $\in \mathbb{R}^{T\times N\times C}$. $U^{\text{reshape}}$, $Z^{\text{reshape}}$, $V$ and $Z_{\text{out}}$ $\in \mathbb{R}^{C \times N\times T}$.  $V^{\text{reshape}}$ and $Z$ $ \in \mathbb{R}^{N\times T \times C}$. $f(\cdot)$ is the mapping function defined by the F-LSTM, which is always bi-directional LSTM and is applied to the frequency dimension of $H_r^{\text{reshape}}$. $h(\cdot)$ is the mapping function defined by the T-LSTM, which scans the time axis. The complex decoder is followed by the Deepfilter\cite{mack2019deep} with looking forward one frame, and the 2 complex LSTM layers defined in~\cite{hu2020dccrn} is finally used to estimate the complex mask for $y(n)$.

The detailed description of our model configuration is shown in Table~\ref{config}. The complex Conv2d/Transpose-Conv2d layers' hyperparameters are given in (kernel size, strides, out channels) format. We omit the Dense layer after each LSTM which keeps the dimension consistent with the input tensor.

\begin{table}[]
\footnotesize
\setlength{\tabcolsep}{3pt}
\caption{Configuration of our proposed method. c- stands for the abbreviation of complex. $\times \text{2}$ means real and imaginary part of a complex kernel.}
\label{config}
\centering
\begin{tabular}[htb]{|l|l|c|l|}
\hline
\multicolumn{1}{|c|}{{Layer name}} & \multicolumn{1}{|c|}{{Input size}} & {\begin{tabular}[|c|]{@{}c@{}}Hyperparameters\end{tabular}} & \multicolumn{1}{|c|}{{Output size}} \\
\hline
c-conv2d\_1 ($\times \text{2}$) & $\text{4} \times \text{T} \times \text{161}$ & $\text{5} \times \text{1, (1, 2), 64}$ & $\text{64} \times \text{T} \times \text{79}$ \\ \hline
c-conv2d\_2 ($\times \text{2}$) & $\text{64} \times \text{T} \times \text{79}$ & $\text{3} \times \text{1, (1, 1), 192}$ & $\text{192} \times \text{T} \times \text{79}$ \\ \hline
reshape\_1 & $\text{192} \times \text{T} \times \text{79}$ & - & $\text{T} \times \text{79} \times \text{192}$ \\ \hline
F-LSTM & $\text{T} \times \text{79} \times \text{96}$ & $\text{128}$ & $\text{T} \times \text{79} \times \text{96}$ \\ \hline
reshape\_2 & $\text{T} \times \text{79} \times \text{96}$ & - & $\text{79} \times \text{T} \times \text{96}$ \\ \hline
T-LSTM & $\text{79} \times \text{T} \times \text{96}$ & $\text{128}$ & $\text{79} \times \text{T} \times \text{96}$ \\ \hline
c-deconv2d\_2 ($\times \text{2}$)& $\text{192} \times \text{T} \times \text{79}$ & $\text{3} \times \text{1, (1, 1), 64}$ & $\text{64} \times \text{T} \times \text{79}$ \\ \hline
c-deconv2d\_1 ($\times \text{2}$)& $\text{64} \times \text{T} \times \text{79}$ & $\text{5} \times \text{1, (1, 2), 2}$ & $\text{2} \times \text{T} \times \text{161}$ \\ \hline
Deepfilter & $\text{2} \times \text{T} \times \text{161}$ & $\text{3} \times \text{3, (1, 1), 9}$ & $\text{2} \times \text{T} \times \text{161}$ \\ \hline
c-LSTM ($\times \text{2}$)& $\text{2} \times \text{T} \times \text{161}$ & $\text{128, 2 layers}$ & $\text{2} \times \text{T} \times \text{161}$\\ \hline
\end{tabular}
\vspace{-10pt}
\end{table}

\subsection{Training targets}
\label{ssec:traintarget}

We estimate complex ratio mask (CRM)~\cite{williamson2015complex} optimized by signal approximation (SA). CRM can be defined as:
\setlength\abovedisplayskip{2.0pt}
\setlength\belowdisplayskip{2.0pt}
\begin{equation}
\mathrm{CRM}=\frac{Y_{r} S_{r}+Y_{i} S_{i}}{Y_{r}^{2}+Y_{i}^{2}}+j \frac{Y_{r} S_{i}-Y_{i} S_{r}}{Y_{r}^{2}+Y_{i}^{2}},
\end{equation}
where $Y$ and $S$ denote $y(n)$ and $s(n)$ after STFT respectively.
The final predicted mask of the network $M = M_r + jM_i$ can also be expressed in polar coordinates:
\setlength\abovedisplayskip{2.0pt}
\setlength\belowdisplayskip{2.0pt}
\begin{equation}
\left\{\begin{array}{l}
{M}_{\text {mag}}=\sqrt{{M}_{r}^{2}+{M}_{i}^{2}} \\
{M}_{\text {phase}}=\arctan 2\left({M}_{i}, {M}_{r}\right)
\end{array}\right.,
\end{equation}
and the estimated clean speech $\hat{S}$ can be calculated as below:
\setlength\abovedisplayskip{2.0pt}
\setlength\belowdisplayskip{2.0pt}
\begin{equation}
{S}={Y}_{\mathrm{mag}} \cdot {M}_{\mathrm{mag}} \cdot e^{{Y}_{\text {phase}}+{M}_{\text {phase }}}.
\end{equation}
\subsection{Cost function}
\label{ssec:traingobj}
The cost function is based on SI-SNR~\cite{vincent2006performance}, which has been widely used as an evaluation metric. 
Instead of computing the average SI-SNR loss of the whole utterance, segmented SI-SNR (Seg-SiSNR) split the utterance into different chunks so that it can distinguish the situation of single-talk and double-talk in a sentence. And our experiments prove that Seg-SiSNR works better than SI-SNR in AEC task.
Seg-SiSNR is defined as:
\setlength\abovedisplayskip{2.0pt}
\setlength\belowdisplayskip{2.0pt}
\begin{equation}
\left\{\begin{array}{ll}
s_{\text {target}} & :=(<\hat{s}, s>\cdot s) /\|s\|_{2}^{2} \\
e_{\text {noise}} & :=\hat{s}-s \\
\text {SI-SNR} & :=10 \log 10\left(\frac{\left\|s_{\text {target}}\right\|_{2}^{2}}{\left\|e_{\text {noise}}\right\|_{2}^{2}}\right) \\
\text{Seg-SiSNR} & := \frac{1}{c}\sum_{i=1}^{c}\text{SI-SNR}(\hat{s}_{\text{seg}i}, s_{\text{seg}i})
\end{array}\right.,
\end{equation}
where $s$ and $\hat{s}$ are the clean and estimated time-domain waveform, respectively. $<\cdot, \cdot>$ denotes the dot product between two vectors and $\|\cdot\|_{2}$ is Euclidean norm (L2 norm). $c$ denotes how many chunks are divided from $\hat{s}$ and $s$. $*_{\text{seg}i}$ denotes the $i$-th speech fragment. We calculate Seg-SiSNR loss for $c=1, 10, 20$ and sum them together as the final cost function.
\section{Experiments}
\label{sec:results}

\subsection{Dataset}
\label{ssec:dataset}

We experiment on the AEC-challenge data~\cite{sridhar2020icassp} to validate the proposed method. In order to train the network, four types of signals need to be prepared: near-end speech, background noise, far-end speech and corresponding echo signal.

For near-end speech $s(n)$, the official synthetic dataset contains 10,000 utterances and we select the first 500 utterances as the test set which does not participate in training. The rest 9,500 utterances, together with the 20,000 utterances (about 70 hours) randomly selected from LibriSpeech~\cite{panayotov2015librispeech} train-clean-100 subset are used for training.

For background noise $v(n)$, we randomly select noise from the DNS~\cite{reddy2021interspeech} data (about 80 hours), in which 20,000 of them are used to generate the test set, and the rest is used in training.
    
For far-end speech $x(n)$ and echo signal $d(n)$, similar to the near-end situation, the first 500 sentences of the official synthetic dataset are used as the test set. Besides, we use the real far-end single-talk recordings provided by the AEC-challenge (about 37 hours), which covers a variety of voice devices and echo signal delay.

To make a fair comparison with another competitive method with reproducible codes -- DTLN-AEC~\cite{westhausen2020acoustic}, we also use the data from the AEC-challenge 2020 only for training and testing. To distinguish the results on different data, we use the suffixes *-20 and *-21 to distinguish the datasets used in AEC-challenge 2020 and 2021 respectively.

\subsection{Data augmentation}
\label{ssec:datasyn}

\begin{table*}[!htbp]
\footnotesize
\centering
\caption{In the case of double-talk, we evaluate PESQ and STOI using on-the-fly data generation, SER $\in [-13, 10]$dB, SNR $\in [5, 20]$dB. We evaluate ERLE of the far-end single-talk scenario in the blind test set.}
\vspace{-5pt}
\label{metric1}
\begin{tabular}{ccccccc}
\toprule
 &  & \multicolumn{2}{c}{\textbf{Clean}}  & \multicolumn{2}{c}{\textbf{Noisy}} & \multicolumn{1}{c}{\textbf{Blind test}}\\
\midrule
\textbf{Method}  &
\textbf{\#Params (M)}   &
\textbf{PESQ}  &
\textbf{STOI}  &
\textbf{PESQ}  &
\textbf{STOI} &
\textbf{ERLE} \\ 
\midrule
Orig  & - & 1.31   & 0.83   & 1.26  & 0.82 & - \cr
\midrule
\begin{tabular}[c]{@{}l@{}}WebRTC-AEC3 \cr
BLSTM-20\cr
Baseline-21\footnotemark \cr 
DTLN-AEC-20\footnotemark \cr
DC-C-T-LSTM-CLSTM-20\end{tabular} 
& \begin{tabular}[c]{@{}l@{}}  - \cr8.09\cr 1.30 \cr 10.42\cr1.35\end{tabular} 
& \begin{tabular}[c]{@{}l@{}} 1.16\cr 1.64\cr 1.66 \cr 2.04\cr1.80\end{tabular} 
& \begin{tabular}[c]{@{}l@{}} 0.63 \cr0.88\cr 0.89\cr 0.58\cr 0.90\end{tabular} 
& \begin{tabular}[c]{@{}l@{}} 1.16\cr1.51\cr 1.47 \cr  1.88\cr 1.68\end{tabular} 
& \begin{tabular}[c]{@{}l@{}} 0.61\cr0.87\cr 0.88\cr  0.57\cr 0.89\end{tabular}
& \begin{tabular}[c]{@{}l@{}} 17.40\cr 21.29\cr 18.97\cr 31.99\cr 23.98\end{tabular}\cr
\midrule
\begin{tabular}[c]{@{}l@{}}DC-F-T-LSTM-CLSTM-20\cr + Dataset-21\cr + Seg-SiSNR\end{tabular}
& \begin{tabular}[c]{@{}l@{}}1.41\end{tabular}
& \begin{tabular}[c]{@{}l@{}}2.12\cr 2.08\cr \textbf{2.13}\end{tabular}    
& \begin{tabular}[c]{@{}l@{}}0.93\cr0.93\cr \textbf{0.93}\end{tabular}   
& \begin{tabular}[c]{@{}l@{}}1.94\cr 1.91\cr \textbf{1.95}\end{tabular}   
& \begin{tabular}[c]{@{}l@{}}0.92\cr 0.92\cr \textbf{0.92}\end{tabular}   
& \begin{tabular}[c]{@{}l@{}} 30.43\cr 30.06\cr \textbf{33.39}\end{tabular}\cr
\bottomrule
\vspace{-15pt}
\end{tabular}
\end{table*}

\begin{figure*}[htb]
\begin{minipage}[b]{.33\linewidth}
  \centering
  \centerline{\includegraphics[width=5.4cm,height=2cm]{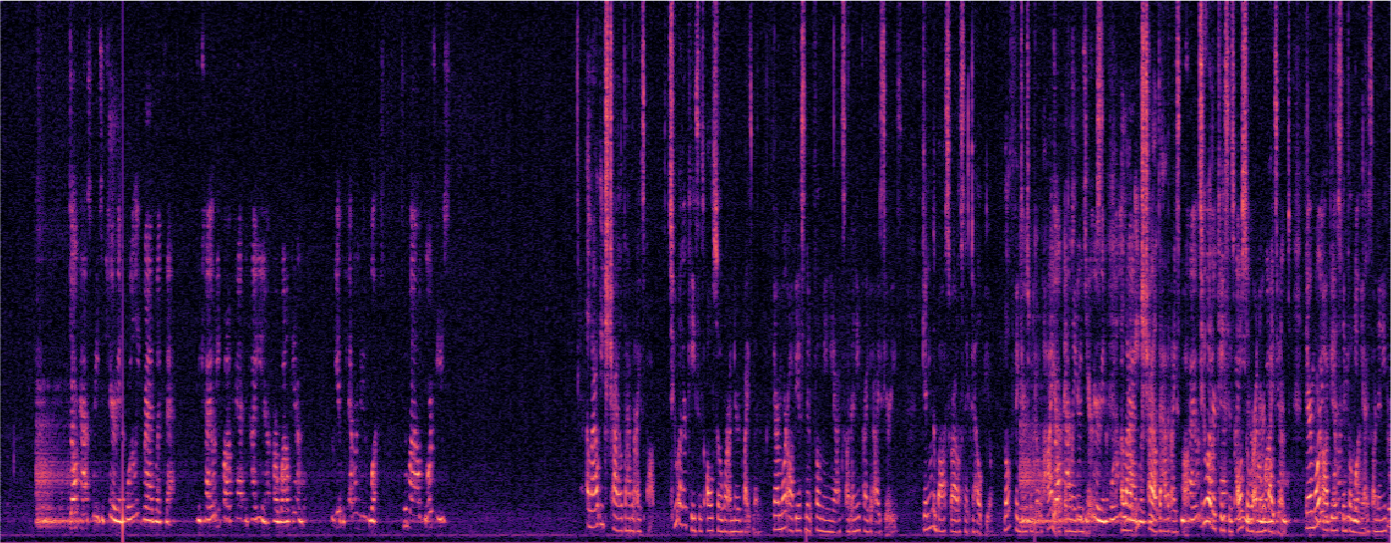}}
  \centerline{(a) Microphone signal}\medskip
  \vspace{-15pt}
\end{minipage}
\begin{minipage}[b]{.33\linewidth}
  \centering
  \centerline{\includegraphics[width=5.4cm,height=2cm]{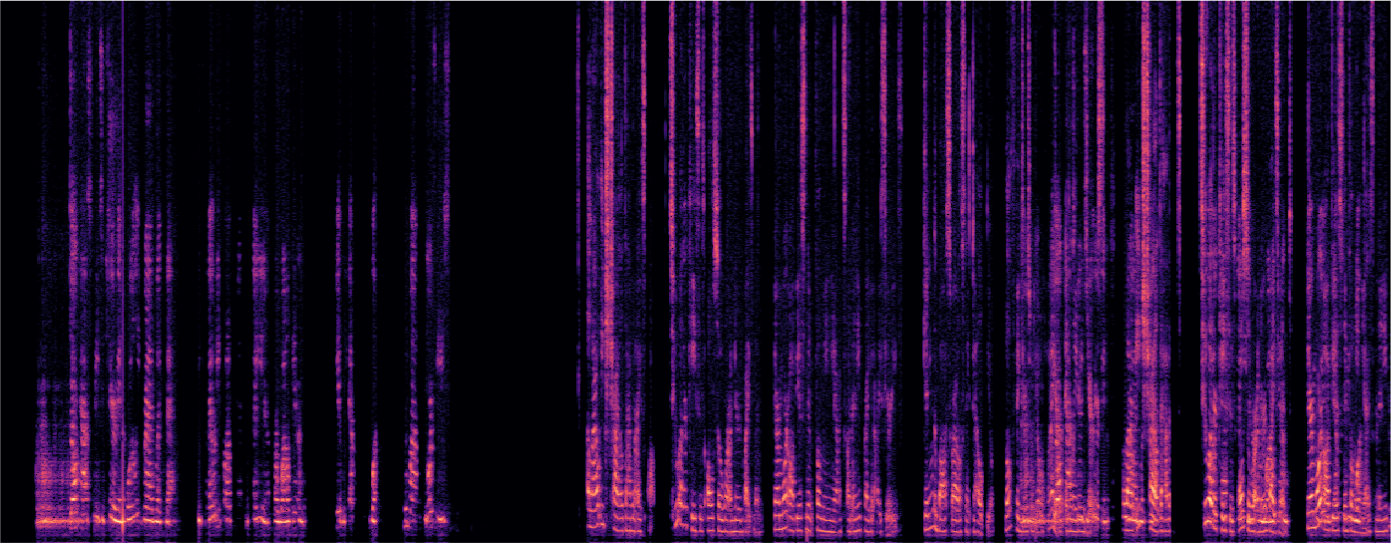}}
  \centerline{(b) Baseline-21}\medskip
  \vspace{-15pt}
\end{minipage}
\begin{minipage}[b]{.33\linewidth}
  \centering
  \centerline{\includegraphics[width=5.4cm,height=2cm]{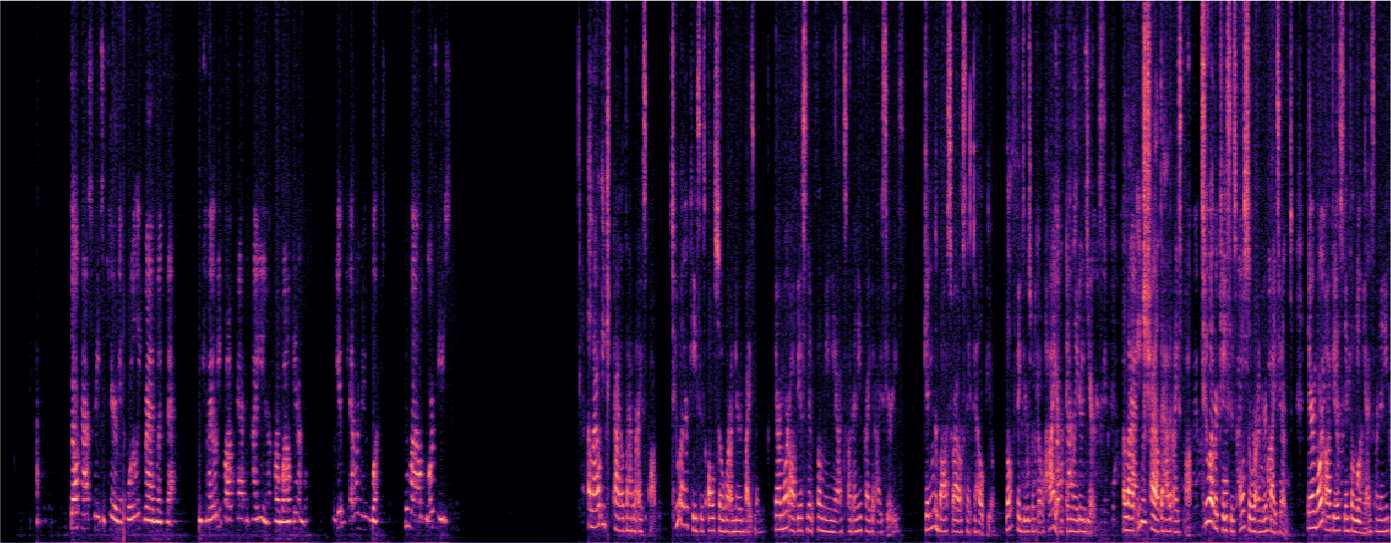}}
  \centerline{(c) DTLN-AEC-20}\medskip
  \vspace{-15pt}
\end{minipage}
\\

\begin{minipage}[b]{.33\linewidth}
  \centering
  \centerline{\includegraphics[width=5.4cm,height=2cm]{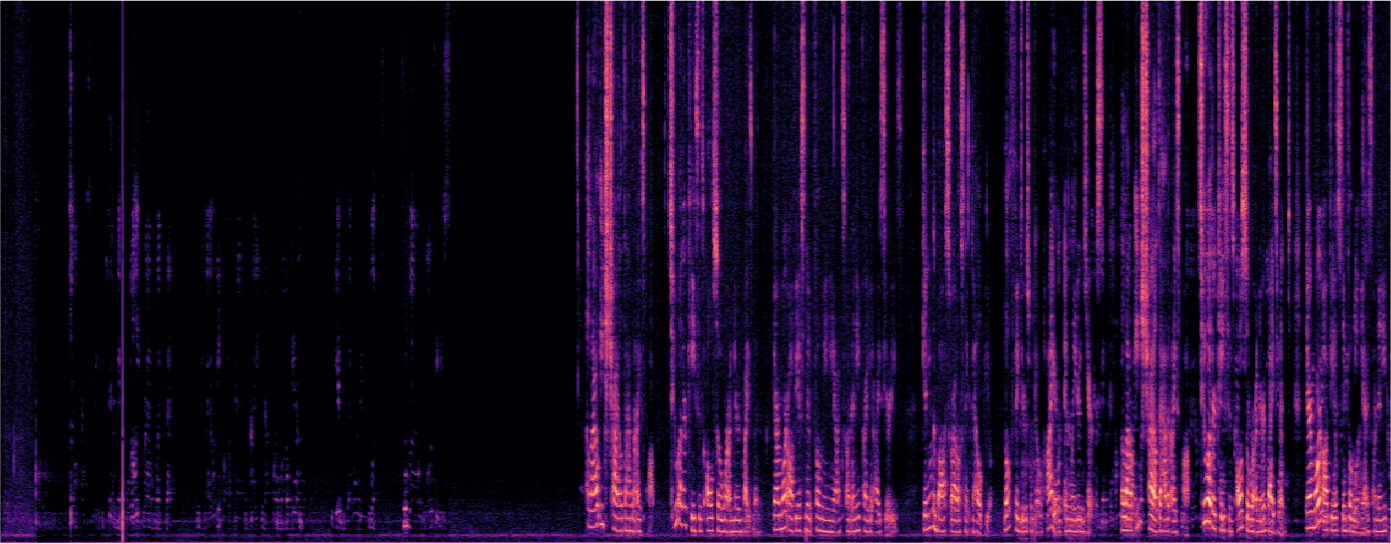}}
  \centerline{(d) DC-F-T-LSTM-CLSTM-20}\medskip
\end{minipage}
\begin{minipage}[b]{.33\linewidth}
  \centering
  \centerline{\includegraphics[width=5.4cm,height=2cm]{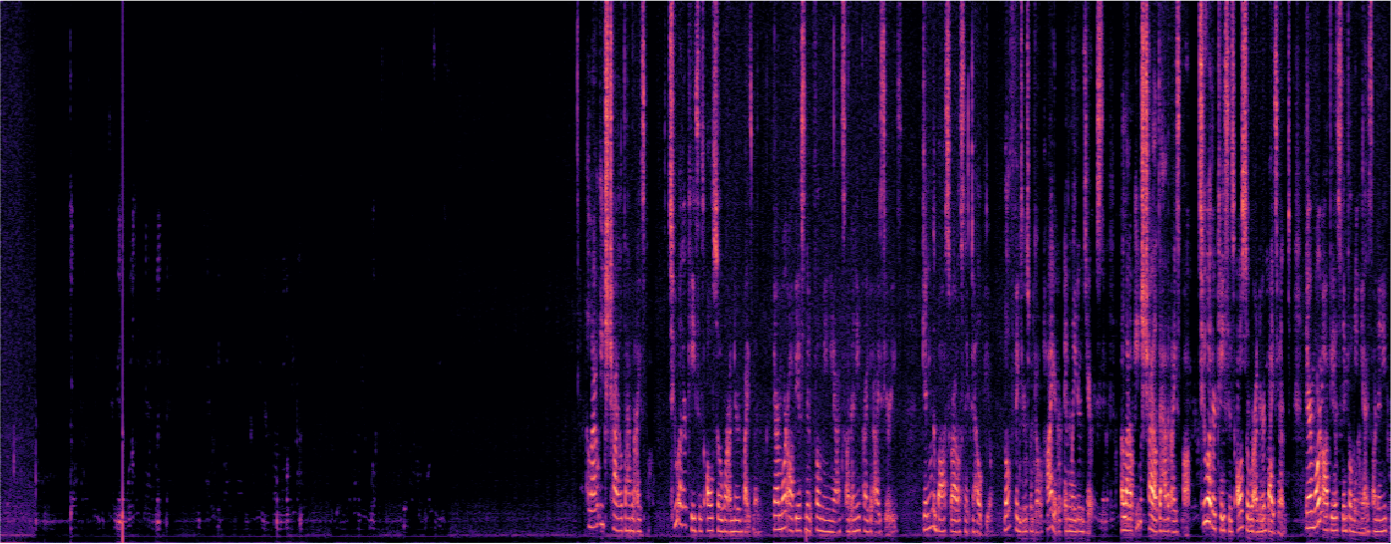}}
  \centerline{(e) DC-F-T-LSTM-CLSTM-21}\medskip
\end{minipage}
\begin{minipage}[b]{.33\linewidth}
  \centering
  \centerline{\includegraphics[width=5.4cm,height=2cm]{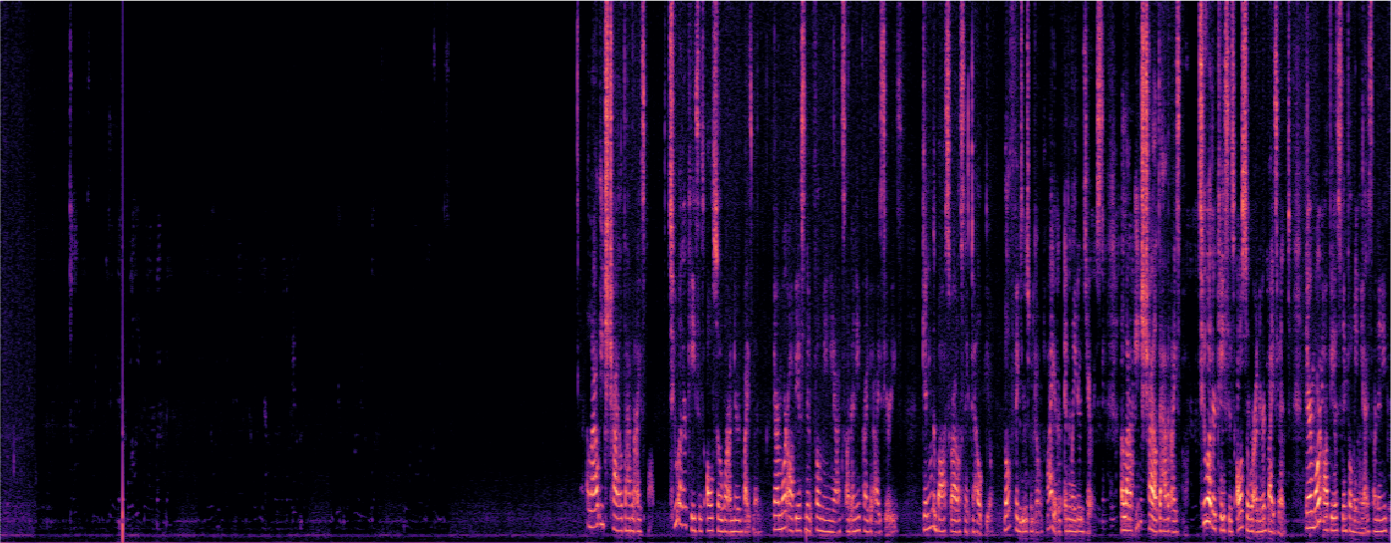}}
  \centerline{(f) DC-F-T-LSTM-CLSTM-21+Seg-SiSNR}\medskip
\end{minipage}
\vspace{-20pt}
\caption{Comparison of different models on real double-talk sample of blind test.}
\label{fig:compare}
\vspace{-15pt}
\end{figure*}

\textbf{Online data generation}. we prepare the near-end speech $s(n)$, background noise $v(n)$, far-end speech $x(n)$ and echo signal $d(n)$ before training, and combine these four signals according to randomly selected signal-to-noise ratio (SNR), signal-to-echo ratio (SER) or other probability factors. In our implement, $\text{SNR} \in[5, 20]$dB and $\text{SER}\in[-10,13]$dB.  The SNR and SER, which are evaluated during double-talk periods, are defined as:
\setlength\abovedisplayskip{2.0pt}
\setlength\belowdisplayskip{2.0pt}
\begin{equation}
    \text{SNR}=10 \log _{10}\left[\sum_{n} s^{2}(n) / \sum_{n} v^{2}(n)\right]
\end{equation}
and
\setlength\abovedisplayskip{2.0pt}
\setlength\belowdisplayskip{2.0pt}
\begin{equation}
    \text{SER}=10 \log _{10}\left[\sum_{n} s^{2}(n) / \sum_{n} d^{2}(n)\right].
\end{equation}
Other probability factors are set up as follows. There is 30\% probability to set $x(n)$ and $d(n)$ as zeros, so that it can simulate the situation of near-end single-talk, and the noise signal ($v(n)$) is set to 0 with 50\% probability. For on-the-fly data generation, various random factors can ensure the diversity of the training data, especially when the echo signal dataset is insufficient.

\textbf{The delay of the far-end signal}. The far-end signal will undergo various of delays before received by the microphone. As shown in the Fig.~\ref{fig:structure}, this delay cannot be avoided in real conditions. The hardware performance and processing algorithm of the device, as well as network fluctuations during the call, may introduce delays. In the conventional DSP-based method, a time delay estimation (TDE) module is needed to align the microphone and the far-end signal. However, due to non-linear changes and background noise interference, errors easily occur in the TDE estimation in practice. We randomly delay the aligned microphone signal from 0 to 100ms to simulate such kind of errors.
    
\textbf{Gain variations}. We apply a random amplification for the echo signal $d(n)$ and the far-end speech $x(n)$. Specifically, we randomly select 3s segment between $d(n)$ and $x(n)$ to attenuate by 20dB to 30dB. The probability of randomly attenuating the signal is 20\%. In addition, through simple maximum normalization, the amplitude range of $[0.3, 0.9]$ is randomly applied to the two signals, and this variations make the network insensitive to amplitude changes.
    
\textbf{Reverberation for near-end signal}. The room impulse responses (RIRs) are generated using the image method~\cite{allen1979image}. To expand data diversity, we simulate 1,000 different rooms of size $a \times b \times h$m for training mixtures, where $a \in [5, 8]$, $b \in [3, 5]$ and $h \in [3, 4]$. We randomly choose 10 positions in each room with random microphone-loudspeaker (M-L) distance ($[0.5, 5]$m) to generate the RIRs. The length of the RIRs is set to 0.5s and the reverberation time (RT60) is randomly chosen from [0.2, 0.7]s. In total 10,000 RIRs are created. We use the first 500 RIRs to generate the test set, and the rest is used for training. For on-the-fly data generation, RIRs are only used to convolve with near-end speech $s(n)$ with 50\% probability. The far-end speech $x(n)$ and echo signal $d(n)$ are either already reverberated or real recordings in different rooms~\cite{sridhar2020icassp}, so there is no need for reverberation.

\begin{table}[]
\setlength{\tabcolsep}{5pt}
\footnotesize
\caption{Subjective ratings in terms of MOS for the blind test set of the AEC-challenge. The confidence interval is 0.02 (ST = single-talk, DT = double talk, NE = near-end, FE = far-end, DT-ECHO means more associated with residual echo, DT-Other means more related to other degradations).}
\vspace{-1pt}
\label{metric2}
\centering
\begin{tabular}[htb]{lccccc}
\toprule
\multicolumn{1}{c}{\textbf{Method}} & \textbf{\begin{tabular}[c]{@{}c@{}}ST-NE\\ MOS\end{tabular}} & \textbf{\begin{tabular}[c]{@{}c@{}}ST-FE\\ DMOS\end{tabular}} & \textbf{\begin{tabular}[c]{@{}c@{}}DT-ECHO\\ DMOS\end{tabular}} & \textbf{\begin{tabular}[c]{@{}c@{}}DT-Other\\ DMOS\end{tabular}}  & \textbf{\begin{tabular}[c]{@{}c@{}}Overall\end{tabular}} \\
\midrule
Baseline  & \textbf{4.18}   & 3.82  & 4.04 & 3.45     & 3.87\\
Ours  & 3.78     & \textbf{4.44} & \textbf{4.44}  & \textbf{3.90}  & \textbf{4.14}\cr
\bottomrule
\end{tabular}
\vspace{-15pt}
\end{table}

\subsection{Performance metrics}

\label{ssec:metrics}
The proposed method is evaluated in terms
of ERLE~\cite{theodoridis2013academic} for single-talk periods. Perceptual evaluation of speech quality (PESQ)~\cite{rix2001perceptual}, short-time objective intelligibility (STOI)~\cite{taal2010short} are used for double-talk periods. The AEC-challenge also provides subjective evaluation results based on the average P.808 Mean Opinion Score (MOS)~\cite{naderi2020open}.
In this study, ERLE is defined as:
\setlength\abovedisplayskip{2.0pt}
\setlength\belowdisplayskip{2.0pt}
\begin{equation}
\mathrm{ERLE}=10\log_{10}\left[\sum_{n} y^{2}(n) / \sum_{n} \hat{s}^{2}(n)\right].
\end{equation}
This variant of ERLE reflects the integrated echo and noise attenuation achieved by system, which is closer to the actual application scenario.

\subsection{Experimental setup}
\label{ssec:expsettings}
Window length and hop size are 20ms and 10ms. Then a 320-point short-time Fourier transform (STFT) is applied to each time frame to produce the complex spectra.
Chunk size of our training data is set to 10s. Our model is trained with the Adam optimizer~\cite{kingma2014adam} for 100 epochs with an initial learning rate of 1e-3, and the learning rate needs to be halved if there is no improvement for two epochs. The whole parameters of model are 1.4M, using SI-SNR loss for training or Seg-SiSNR loss if specially pointed out. Overall delay of the system is 40ms. The real time factor (RTF) of our network is 0.4385, tested on Intel(R) Xeon(R) CPU E5-2640@2.50GHz with single-core. Some of the processed audio clips can be found in this page\footnotemark.

\footnotetext[1]{https://github.com/microsoft/AEC-challenge}
\footnotetext[2]{https://github.com/breizhn/DTLN-aec}
\footnotetext[3]{https://echocatzh.github.io/Demo-of-DeepComplexAEC}

\subsection{Results and Analysis}
\label{ssec:performace}

In Table~\ref{metric1}, we compare different methods in the AEC-challenge datasets.
Because of non-linear distortion and noise interference, WebRTC-AEC3 does not work well in the view of PESQ and STOI. Our method outperforms the BLSTM~\cite{zhang2018deep} (4 BLSTM layers with 300 hidden units) and AEC-challenge baseline~\cite{sridhar2020icassp} (2 GRU layers with 322 hidden units) in all conditions.
Besides the DC-F-T-LSTM-CLSTM which recurrents in frequency as well as time axis, we tried DC-C-T-LSTM-CLSTM which recurrents in channel and time axis for comparison. With almost the same amount of parameters, our experiment proves that doing recurrence in the frequency axis is more effective.  Compared with DTLN-AEC-20 , DC-F-T-LSTM-CLSTM-20 clearly brings better performance with less parameters. 
Dataset-21 means AEC-challenge 2021 dataset and 60h near-end speech from LibriSpeech. We notice that PESQ gets worse with the use of more real data, this is because some invalid far-end single-talk clips, which contain near-end speech, are not eliminated. Even with these invalid clips, using Seg-SiSNR as cost function shows improvement and achieves the best results. Fig.~\ref{fig:compare} exhibits the improvement of our methods under the same training dataset and better ability to suppress residual noise using Seg-SiSNR loss. 

Table~\ref{metric2} shows that our method significantly outperforms the AEC-challenge baseline except for the ST-NE condition. The overall MOS improvement is as high as 0.27. The ST-NE case may be caused by narrow range of SER ($[-13, 10]$ dB) and SNR($[5, 20]$ dB) when generating the training data on-the-fly, resulting in insufficient data coverage of ST-NE scenario (high SNR/SER scenario) and leading to perceivable speech distortion in this scenario. We will solve this problem in the future.

\section{Conclusions}
This study has shown that our proposed neural AEC system  -- DC-F-T-LSTM-CLSTM, with smaller size of parameters and lower runtime delay, can achieve better echo cancellation and noise suppression performance compared with the competitive methods. We verify that the magnitude and phase information can be more effectively used with the complex operation and the F-T-LSTM module. With Seg-SiSNR as the cost function, the residual echo and noise can be suppressed further. Experimental results in double-talk scenarios, background noise situations and real recordings were also reported, proving that our method is effective in challenging acoustic echo conditions
In future work, we will optimize the data generation strategy for adapting to the real acoustic environment better, and consider lower complexity and hybrid DSP/Neural network methods.

\bibliographystyle{IEEEtran}

\bibliography{main}

\begin{thebibliography}{10}
\providecommand{\url}[1]{#1}
\csname url@samestyle\endcsname
\providecommand{\newblock}{\relax}
\providecommand{\bibinfo}[2]{#2}
\providecommand{\BIBentrySTDinterwordspacing}{\spaceskip=0pt\relax}
\providecommand{\BIBentryALTinterwordstretchfactor}{4}
\providecommand{\BIBentryALTinterwordspacing}{\spaceskip=\fontdimen2\font plus
\BIBentryALTinterwordstretchfactor\fontdimen3\font minus
  \fontdimen4\font\relax}
\providecommand{\BIBforeignlanguage}[2]{{%
\expandafter\ifx\csname l@#1\endcsname\relax
\typeout{** WARNING: IEEEtran.bst: No hyphenation pattern has been}%
\typeout{** loaded for the language `#1'. Using the pattern for}%
\typeout{** the default language instead.}%
\else
\language=\csname l@#1\endcsname
\fi
#2}}
\providecommand{\BIBdecl}{\relax}
\BIBdecl

\bibitem{benesty2007springer}
J.~Benesty, M.~M. Sondhi, and Y.~Huang, \emph{Springer handbook of speech
  processing}.\hskip 1em plus 0.5em minus 0.4em\relax Springer, 2007.

\bibitem{1163949}
D.~Mansour and A.~Gray, ``Unconstrained frequency-domain adaptive filter,''
  \emph{IEEE Transactions on Acoustics, Speech, and Signal Processing},
  vol.~30, no.~5, pp. 726--734, 1982.

\bibitem{103078}
J.-S. Soo and K.~Pang, ``Multidelay block frequency domain adaptive filter,''
  \emph{IEEE Transactions on Acoustics, Speech, and Signal Processing},
  vol.~38, no.~2, pp. 373--376, 1990.

\bibitem{2002A}
S.~Gustafsson, R.~Martin, P.~Jax, and P.~Vary, ``A psychoacoustic approach to
  combined acoustic echo cancellation and noise reduction,'' \emph{IEEE
  Transactions on Speech and Audio Processing}, vol.~10, no.~5, pp. 245--256,
  2002.

\bibitem{4517596}
D.~A. Bendersky, J.~W. Stokes, and H.~S. Malvar, ``Nonlinear residual acoustic
  echo suppression for high levels of harmonic distortion,'' in \emph{2008 IEEE
  International Conference on Acoustics, Speech and Signal Processing}, 2008,
  pp. 261--264.

\bibitem{hansler2005acoustic}
E.~H{\"a}nsler and G.~Schmidt, \emph{Acoustic echo and noise control: a
  practical approach}.\hskip 1em plus 0.5em minus 0.4em\relax John Wiley and
  Sons, 2005, vol.~40.

\bibitem{turbin1997comparison}
V.~Turbin, A.~Gilloire, and P.~Scalart, ``Comparison of three post-filtering
  algorithms for residual acoustic echo reduction,'' in \emph{1997 IEEE
  International Conference on Acoustics, Speech, and Signal Processing},
  vol.~1, 1997, pp. 307--310 vol.1.

\bibitem{boll1979suppression}
S.~Boll, ``Suppression of acoustic noise in speech using spectral
  subtraction,'' \emph{IEEE Transactions on acoustics, speech, and signal
  processing}, vol.~27, no.~2, pp. 113--120, 1979.

\bibitem{ma2020acoustic}
L.~Ma, H.~Huang, P.~Zhao, and T.~Su, ``Acoustic echo cancellation by combining
  adaptive digital filter and recurrent neural network,'' 2020.

\bibitem{fazel2020cad}
A.~Fazel, M.~El-Khamy, and J.~Lee, ``Cad-aec: Context-aware deep acoustic echo
  cancellation,'' in \emph{ICASSP 2020 - 2020 IEEE International Conference on
  Acoustics, Speech and Signal Processing (ICASSP)}, 2020, pp. 6919--6923.

\bibitem{wang2021weighted}
Z.~Wang, Y.~Na, Z.~Liu, B.~Tian, and Q.~Fu, ``Weighted recursive least square
  filter and neural network based residual echo suppression for the
  aec-challenge,'' in \emph{ICASSP 2021 - 2021 IEEE International Conference on
  Acoustics, Speech and Signal Processing (ICASSP)}, 2021, pp. 141--145.

\bibitem{valin2021lowcomplexity}
J.-M. Valin, S.~Tenneti, K.~Helwani, U.~Isik, and A.~Krishnaswamy,
  ``Low-complexity, real-time joint neural echo control and speech enhancement
  based on percepnet,'' in \emph{ICASSP 2021 - 2021 IEEE International
  Conference on Acoustics, Speech and Signal Processing (ICASSP)}, 2021, pp.
  7133--7137.

\bibitem{sridhar2020icassp}
R.~Cutler, A.~Saabas, T.~Parnamaa, M.~Loide, S.~Sootla, M.~Purin, H.~Gamper,
  S.~Braun, K.~Sorensen, R.~Aichner, and S.~Srinivasan, ``Interspeech 2021
  acoustic echo cancellation challenge: Datasets and testing framework,'' in
  \emph{INTERSPEECH 2021}, 2021.

\bibitem{zhang2018deep}
H.~Zhang and D.~Wang, ``Deep learning for acoustic echo cancellation in noisy
  and double-talk scenarios,'' \emph{Training}, vol. 161, no.~2, p. 322, 2018.

\bibitem{westhausen2020acoustic}
W.~N. L. and M.~B. T., ``Acoustic echo cancellation with the dual-signal
  transformation lstm network,'' in \emph{ICASSP 2021 - 2021 IEEE International
  Conference on Acoustics, Speech and Signal Processing (ICASSP)}, 2021, pp.
  7138--7142.

\bibitem{westhausen2020dual}
N.~L. Westhausen and B.~T. Meyer, ``Dual-signal transformation lstm network for
  real-time noise suppression,'' \emph{arXiv preprint arXiv:2005.07551}, 2020.

\bibitem{chen2020nonlinear}
H.~Chen, T.~Xiang, K.~Chen, and J.~Lu, ``Nonlinear residual echo suppression
  based on multi-stream conv-tasnet,'' 2020.

\bibitem{Luo_2019}
Y.~Luo and N.~Mesgarani, ``Conv-tasnet: Surpassing ideal time–frequency
  magnitude masking for speech separation,'' \emph{IEEE/ACM Transactions on
  Audio, Speech, and Language Processing}, vol.~27, no.~8, pp. 1256--1266,
  2019.

\bibitem{kim2020attention}
J.-H. Kim and J.-H. Chang, ``Attention wave-u-net for acoustic echo
  cancellation,'' \emph{Proc. Interspeech 2020}, pp. 3969--3973, 2020.

\bibitem{stoller2018wave}
D.~Stoller, S.~Ewert, and S.~Dixon, ``Wave-u-net: A multi-scale neural network
  for end-to-end audio source separation,'' \emph{arXiv preprint
  arXiv:1806.03185}, 2018.

\bibitem{choi2018phase}
H.-S. Choi, J.-H. Kim, J.~Huh, A.~Kim, J.-W. Ha, and K.~Lee, ``Phase-aware
  speech enhancement with deep complex u-net,'' \emph{arXiv e-prints}, pp.
  arXiv--1903, 2019.

\bibitem{hu2020dccrn}
Y.~Hu, Y.~Liu, S.~Lv, M.~Xing, S.~Zhang, Y.~Fu, J.~Wu, B.~Zhang, and L.~Xie,
  ``Dccrn: Deep complex convolution recurrent network for phase-aware speech
  enhancement,'' \emph{arXiv preprint arXiv:2008.00264}, 2020.

\bibitem{li2015lstm}
J.~Li, A.~Mohamed, G.~Zweig, and Y.~Gong, ``Lstm time and frequency recurrence
  for automatic speech recognition,'' in \emph{2015 IEEE Workshop on Automatic
  Speech Recognition and Understanding (ASRU)}, 2015, pp. 187--191.

\bibitem{mack2019deep}
W.~Mack and E.~A.~P. Habets, ``Deep filtering: Signal extraction and
  reconstruction using complex time-frequency filters,'' \emph{IEEE Signal
  Processing Letters}, vol.~27, pp. 61--65, 2020.

\bibitem{williamson2015complex}
D.~S. Williamson, Y.~Wang, and D.~Wang, ``Complex ratio masking for monaural
  speech separation,'' \emph{IEEE/ACM Transactions on Audio, Speech, and
  Language Processing}, vol.~24, no.~3, pp. 483--492, 2016.

\bibitem{vincent2006performance}
E.~Vincent, R.~Gribonval, and C.~Fevotte, ``Performance measurement in blind
  audio source separation,'' \emph{IEEE Transactions on Audio, Speech, and
  Language Processing}, vol.~14, no.~4, pp. 1462--1469, 2006.

\bibitem{panayotov2015librispeech}
V.~Panayotov, G.~Chen, D.~Povey, and S.~Khudanpur, ``Librispeech: An asr corpus
  based on public domain audio books,'' in \emph{2015 IEEE International
  Conference on Acoustics, Speech and Signal Processing (ICASSP)}, 2015, pp.
  5206--5210.

\bibitem{reddy2021interspeech}
C.~K. Reddy, H.~Dubey, K.~Koishida, A.~Nair, V.~Gopal, R.~Cutler, S.~Braun,
  H.~Gamper, R.~Aichner, and S.~Srinivasan, ``Interspeech 2021 deep noise
  suppression challenge,'' \emph{arXiv preprint arXiv:2101.01902}, 2021.

\bibitem{allen1979image}
J.~B. Allen and D.~A. Berkley, ``Image method for efficiently simulating
  small-room acoustics,'' \emph{The Journal of the Acoustical Society of
  America}, vol.~65, no.~4, pp. 943--950, 1979.

\bibitem{theodoridis2013academic}
S.~Theodoridis and R.~Chellappa, \emph{Academic Press Library in Signal
  Processing: Image, Video Processing and Analysis, Hardware, Audio, Acoustic
  and Speech Processing}.\hskip 1em plus 0.5em minus 0.4em\relax Academic
  Press, 2013.

\bibitem{rix2001perceptual}
A.~W. Rix, J.~G. Beerends, M.~P. Hollier, and A.~P. Hekstra, ``Perceptual
  evaluation of speech quality (pesq)-a new method for speech quality
  assessment of telephone networks and codecs,'' in \emph{2001 IEEE
  International Conference on Acoustics, Speech, and Signal Processing.
  Proceedings (Cat. No. 01CH37221)}, vol.~2.\hskip 1em plus 0.5em minus
  0.4em\relax IEEE, 2001, pp. 749--752.

\bibitem{taal2010short}
C.~H. Taal, R.~C. Hendriks, R.~Heusdens, and J.~Jensen, ``A short-time
  objective intelligibility measure for time-frequency weighted noisy speech,''
  in \emph{2010 IEEE International Conference on Acoustics, Speech and Signal
  Processing}, 2010, pp. 4214--4217.

\bibitem{naderi2020open}
R.~Cutler, B.~Nadari, M.~Loide, S.~Sootla, and A.~Saabas, ``Crowdsourcing
  approach for subjective evaluation of echo impairment,'' in \emph{ICASSP
  2021-2021 IEEE International Conference on Acoustics, Speech and Signal
  Processing (ICASSP)}.\hskip 1em plus 0.5em minus 0.4em\relax IEEE, 2021, pp.
  406--410.

\bibitem{kingma2014adam}
D.~P. Kingma and J.~Ba, ``Adam: A method for stochastic optimization,''
  \emph{arXiv preprint arXiv:1412.6980}, 2014.

\end{thebibliography}

\end{document}